\documentclass[twocolumn]{aastex62}
\pdfoutput=1

\usepackage{amsmath,graphicx,multirow}

\received{}
\revised{}
\accepted{}
\submitjournal{ApJ}

\shorttitle{Interpreting Star Count Ratios in Stellar Populations}
\shortauthors{Dorn-Wallenstein \& Levesque}

\begin{document}

\title{A Comparison of Rotating and Binary Stellar Evolution Models: Effects on Massive Star Populations}

\correspondingauthor{Trevor Z. Dorn-Wallenstein}
\email{tzdw@uw.edu}

\author[0000-0003-3601-3180]{Trevor Z. Dorn-Wallenstein}
\affiliation{University of Washington Astronomy Department \\
Physics and Astronomy Building, 3910 15th Ave NE  \\
Seattle, WA 98105, USA} 

\author[0000-0003-2184-1581]{Emily M. Levesque}
\affiliation{University of Washington Astronomy Department \\
Physics and Astronomy Building, 3910 15th Ave NE  \\
Seattle, WA 98105, USA}

\begin{abstract}
Both rotation and interactions with binary companions can significantly affect massive star evolution, altering interior and surface abundances, mass loss rates and mechanisms, observed temperatures and luminosities, and their ultimate core-collapse fates. The Geneva and BPASS stellar evolution codes include detailed treatments of rotation and binary evolutionary effects, respectively, and can illustrate the impact of these phenomena on massive stars and stellar populations. However, a direct comparison of these two widely-used codes is vital if we hope to use their predictions for interpreting observations. In particular, rotating and binary models will predict different young stellar populations, impacting the outputs of stellar population synthesis (SPS) and the resulting interpretation of large massive star samples based on commonly-used tools such as star count ratios. Here we compare the Geneva and BPASS evolutionary models, using an interpolated SPS scheme introduced in our previous work and a novel Bayesian framework to present the first in-depth direct comparison of massive stellar populations produced from single, rotating, and binary non-rotating evolution models. We calculate both models' predicted values of star count ratios and compare the results to observations of massive stars in Westerlund 1, $h + \chi$ Persei, and both Magellanic Clouds. We also consider the limitations of both the observations and the models, and how to quantitatively include observational completeness limits in SPS models. We demonstrate that the methods presented here, when combined with robust stellar evolutionary models, offer a potential means of estimating the physical properties of massive stars in large stellar populations.

\end{abstract}


\section{Introduction} \label{sec:intro}

Rotation is a ubiquitous property of stars, and has a significant effect on the physical properties of massive ($M_{ini}\gtrsim8 M_\odot$) stars. Rotationally induced mixing via diffusion and meridional circulation \citep{zahn92} alters stellar interiors enhances surface element abundances, centrifugal forces alter stellar shapes \citep{vonzeipel24}, and both radiative \citep{maeder00} and mechanical \citep{georgy10} mass loss are boosted by rotation. The Geneva code \citep{ekstrom12,georgy13,groh19} is the current state of the art implementation of rotating stellar evolutionary models, and has been used to study the impact of rotation on the radiative output \citep{levesque12}, chemical yields \citep{hirschi05}, and final fates \citep{meynet15} of massive stars.

Simultaneously, many massive stars are born into binary and higher-order systems \citep{sana12,duchene13,sana14,moe17}, and binary systems with dwarf, supergiant, Wolf-Rayet, and compact components are commonly observed \citep{sana13,neugent14,neugent18b,neugent19}. Interactions in binary systems through both tides \citep{hurley02} and mass transfer can completely disrupt the evolution of both stars, creating otherwise-impossible evolutionary states. Both the detailed makeups and integrated spectral energy distributions of populations of massive stars are affected, especially at low metallicity \citep{stanway16}. The Binary Population and Spectral Synthesis code (BPASS, \citealt{eldridge17,stanway18}) is the current state of the art rigorous implementation of these effects. 

Despite the importance of binary interactions and rotation, the theory of single, nonrotating stars has thusfar been successful at predicting the evolution of stars of varying compositions and initial masses (the ``Conti scenario'', \citealt{conti83}). The addition of the effects of rotation has further refined single-star evolution to reproduce observed surface abundance enhancements and mass loss rates \citep[e.g.,][]{meynet00}. These successes are often used as evidence that binary interactions are a secondary effect, or can be approximated by simple methods \citep[e.g., by instantaneously removing a model star's H envelope after it leaves the main sequence; for a review of recent stellar evolution models featuring simple implementations of binary interactions, see][and citations therein]{eldridge17}. Indeed, at first glance some of the effects predicted by binary stellar evolution models (such as enhanced surface abundances, stronger mass loss, broader mass and age ranges for Wolf-Rayet progenitors, and harder ionizing spectra produced by stellar populations) are identical to the predictions from rotating stellar models (e.g. \citealt{levesque12}). Regardless, massive binary systems {\it do} exist, necessitating detailed modeling of their evolution as well as observations of the frequency of binary systems to understand their impact on stellar populations. 

Finding binary systems via radial velocity variability (e.g., \citealt{sana12}; \citealt{sana13}) requires long time baselines, high spectral resolution, and long integration times, and is currently infeasible for stars beyond the Magellanic Clouds (MCs), or for some stars such as red supergiant binaries \citep{neugent19}. For more-distant systems, we are left studying (semi-)resolved stellar populations. Simultaneously, measuring rotation periods requires high-cadence time-series photometry \citep{blomme11,buysschaert15,balona15,balona16,johnston17,ramiaramanantsoa18,pedersen19}. Spectroscopic measurements yield projected rotational velocities, with the caveat that the inclination of the rotational axis to the line of sight is unknown \citep[e.g., ][]{huang10}. In either case, both methods again require high quality observations of nearby stars. Fortunately, massive stars are luminous, and can be resolved in galaxies around the Local Group and beyond. Previously, we used the BPASS models to construct grids of synthetic populations with varying metallicity ($Z$), star formation history (SFH), and the natal binary fraction ($f_{bin}$), and predicted the frequency of various evolutionary phases \citep[][Paper I hereafter]{dornwallenstein18}. Here, we incorporate stellar evolution models that include rotation \citep{ekstrom12,georgy13,groh19}, and perform a detailed comparison between the two model sets.

\begin{deluxetable*}{lll}
\tabletypesize{\normalsize}
\tablecaption{Model parameters used to label model timesteps with an evolutionary phase, and other variables introduced in the text.\label{tab:mod_params}}
\tablehead{\colhead{Parameter} &
\colhead{Description} & \colhead{Unit}} 
\startdata
$\log(L)$ & Logarithm of the luminosity & $L_\odot$ \\
$\log(T_{eff})$ & Logarithm of the effective temperature & K \\
$\log(g)$ & Logarithm of the surface gravity & cm s$^{-2}$ \\
$X$ & Hydrogen Surface Mass Fraction & - \\
$Y$ & Helium Surface Mass Fraction & - \\
$C$ & Carbon Surface Mass Fraction (Sum of $^{12}$C and $^{13}$C for Geneva tracks) & - \\
$N$ & Nitrogen Surface Mass Fraction & - \\
$O$ & Oxygen Surface Mass Fraction & - \\
$\log{t}$ & Logarithm of time & yr \\
$Z$ & Mass fraction metals, $Z_\odot = 0.014$ & - \\
$f_{bin}$ & Binary fraction & - \\
$f_{rot}$ & Rotating fraction & - \\
$f$ & Generic term for either $f_{bin}$ or $f_{rot}$ & - \\
$n_S$ & Observed frequency of an arbitrary spectral type S & - \\
$R_{S_1/S_2}$ & Observed ratio of the frequency of two spectral types, S$_1$ and S$_2$ & - \\
$\hat{n}_S$ & Intrinsic frequency of an arbitrary spectral type S & - \\
$\hat{R}_{S_1/S_2}$ & Intrinsic ratio of the frequency of two spectral types, defined as $\hat{R}_{S_1/S_2}\equiv\hat{n}_{S_1} / \hat{n}_{S_2}$ & - \\ 
\enddata
\end{deluxetable*}

In \S\ref{sec:method}, we compare the two evolutionary codes, from the BPASS and Geneva groups, used to generate binary and rotating stellar tracks respectively. We also detail the population synthesis method we employ to generate theoretical rotating and binary stellar populations and subsequent predictions for the frequency of various spectral types in these populations. In \S\ref{sec:results} we describe the observables we generate from the two synthetic populations, and introduce a novel Bayesian framework of estimating these observables from data. We compare our synthetic rotating and binary populations to each other and to observations of massive star populations with both simple and complex star-formation histories before considering the results of our comparison and the implications for using rotating and binary stellar evolution models to interpret future observations of massive star populations (\S\ref{sec:conclusion}).

\section{Creating Theoretical Populations}\label{sec:method}
\subsection{The Models}\label{sec:themodels} 

In Paper I, we created synthetic populations using BPASS version 2.2.1, which incorporates the effects of both tides and mass transfer to predict the evolution of single and binary stars on a dense grid of initial primary and secondary masses ($M_1$ and $M_2$), initial periods $P$, and mass ratios ($q\equiv M_2/M1$) at 13 metallicities. We express the metallicity as a mass fraction $Z$, and BPASS adopts metallicities in the range $10^{-5}\leq Z \leq 0.04$. Note that for the duration of this paper, we assume solar metallicity $Z_\odot = 0.014$ \citep{asplund09}. Binary interactions are modeled as enhanced mass loss/gain from/onto its model stars via Roche Lobe Overflow (RLOF). The orbital energy is tracked throughout, allowing BPASS to model a broad range of simulated evolutionary scenarios. 

Here, we also incorporate the Geneva evolutionary tracks. Stellar evolution is modeled at two different initial rotation rates (nonrotating, with $v_{ini}/v_{crit} = 0$, and rotating, with $v_{ini}/v_{crit} = 0.4$) and at three different metallicities: $Z=0.014$ \citep{ekstrom12}, $Z=0.002$ \citep{georgy13}, and $Z=0.0004$ \citep{groh19}. Horizontal diffusion coefficients in the rotating models are calculated following \citet{zahn92}. Meridional circulation is calculated as described by \citet{maeder98}, and the two effects are combined following \citet{chaboyer92}. Angular momentum transport is included, and angular momentum is conserved following \citet{georgy10}. Finally, rotation-enhanced radiative \citep{maeder00} and mechanical \citep{georgy10} mass loss are also implemented; for a detailed disucssion see \citep{ekstrom12}.

\subsection{Population Synthesis}\label{subsec:popsyn}

Evolutionary tracks in hand, we synthesize a population by weighing each track with the initial mass function,  $\Phi(M)$. We adopt the default form of $\Phi$ in BPASS v2.2.1, which is a broken power law with slope -1.3 below 0.5 $M_\odot$, and a slope of -2.35 for higher masses, with a minimum mass of 0.1 $M_\odot$ and a maximum mass of 300 $M_\odot$, normalized so the total stellar mass is $10^6 M_\odot$. Binary models are also weighted according to the distributions of the fundamental natal period $P$ and mass ratio $q$ from \citet{moe17}. Because the Geneva tracks are sampled on a much coarser grid of initial mass than the BPASS single star tracks, we linearly interpolate the available Geneva tracks onto the BPASS single star initial mass grid following \citet{georgy14}, and adopt the IMF weighting from the BPASS v2.2.1 inputs\footnote{Details on the mass grids, parameter values and more can be found in the BPASS v2.2 User Manual, currently hosted online at \url{bpass.auckland.ac.nz}}. Additionally, both rotating and nonrotating tracks at all three metallicities are only available between 1.7 and 120 $M_\odot$. We choose not to introduce any correction factors to the IMF weights --- e.g., boosting the weight of the 120 $M_\odot$ model to represent all stars with initial masses between 120 and 300 $M_\odot$ --- to ensure that tracks with identical masses are weighted identically. Because no single star below the 1.7 $M_\odot$ threshold would become any of the stellar types considered here, only the exclusion of these very massive stars would affect our results. However, these stars are so rare, and their lifetimes so short, that their impact on our synthetic populations is minimal.

We create four sets of synthetic populations: one composed entirely of single, nonrotating stars using the input files provided in the BPASS v2.2 data release ($f_{bin} = 0$), one composed entirely of binary stars ($f_{bin} = 1$) using custom input files provided by the BPASS team (J. J. Eldridge 2018, private communication), one composed of single, nonrotating stars from the Geneva models (hereafter referred to as the population with ``rotating fraction'' $f_{rot} = 0$), and one composed entirely of rotating stars ($f_{rot} = 1$). We note that there are no single stars in the custom $f_{bin}=1$ population, though the distribution of periods and mass ratios is identical to the default BPASS v2.2 binary population, which is drawn from \citet{moe17}.

\subsection{Number Counts vs. Time}\label{subsec:numbers}

\begin{deluxetable*}{lll}
\tabletypesize{\normalsize}
\tablecaption{Criteria used to classify evolution tracks. Adapted from Table 3 of \citet{eldridge17}. We specify where the Geneva tracks are classified with different criteria than the BPASS tracks.\label{tab:count_criteria}}
\tablehead{\colhead{Label} &
\colhead{BPASS Criteria} & \colhead{Geneva Criteria}} 
\startdata
\multirow{2}{*}{WNH} & \multirow{2}{*}{\parbox{4cm}{$\log(T_{eff}) \geq 4.45$ \\ $X \leq 0.4$}} & \multirow{2}{*}{\parbox{4cm}{$\log(T_{eff}) \geq 4.0$ \\ $X \leq 0.3$}}\\
\\
\multirow{3}{*}{WN} & \multirow{3}{*}{\parbox{5cm}{$\log(T_{eff}) \geq 4.45$ \\ $X \leq 10^{-3}$ \\ $(C + O))/Y \leq 0.03$}} & \multirow{3}{*}{\parbox{5cm}{$\log(T_{eff}) \geq 4.0$ \\ $X \leq 10^{-5}$ \\ $N > C$}} \\
\\
\\
\multirow{3}{*}{WC} & \multirow{3}{*}{\parbox{5cm}{$\log(T_{eff}) \geq 4.45$ \\ $X \leq 10^{-3}$ \\ $(C + O))/Y > 0.03$}} & \multirow{3}{*}{\parbox{5cm}{$\log(T_{eff}) \geq 4.0$ \\ $X \leq 10^{-5}$ \\ $N \leq C$}} \\
\\
\\
O & $\log(T_{eff}) \geq 4.48$ & $\log(T_{eff}) \geq 4.5$  \\
\multirow{2}{*}{O$f$} & \multirow{2}{*}{\parbox{5cm}{$\log(T_{eff}) \geq 4.519$ \\ $\log(g) > 3.676 \log(T_{eff}) + 13.253$}} & \\
\\
B & $4.041 \leq \log(T_{eff}) < 4.48$ & $4.041 \leq \log(T_{eff}) < 4.5$ \\
A & $3.9 \leq \log(T_{eff}) < 4.041$ & $3.8 \leq \log(T_{eff}) < 4.041$ \\
F/G & $3.66 \leq \log(T_{eff}) < 3.9$ & $3.66 \leq \log(T_{eff}) < 3.8$ \\
K & $3.55 \leq \log(T_{eff}) < 3.66 $ & \\
M & $\log(T_{eff}) < 3.55$ & \\
\multirow{2}{*}{BSG} & \multirow{2}{*}{\parbox{4cm}{O + O$f$ + B + A \\ $\log(L) \geq 4.9$}} \\
\\
\multirow{2}{*}{YSG} & \multirow{2}{*}{\parbox{4cm}{F/G \\ $\log(L) \geq 4.9$}} \\
\\
\multirow{2}{*}{RSG} & \multirow{2}{*}{\parbox{4cm}{K + M \\ $\log(L) \geq 4.9$}} \\
\\
\multirow{2}{*}{WR} & \multirow{2}{*}{\parbox{4cm}{WNH + WN + WC \\ $\log(L) \geq 4.9$}} \\
\\
\enddata
\end{deluxetable*}

\begin{figure*}[p!]
\centering
\includegraphics[scale=0.8,angle=90]{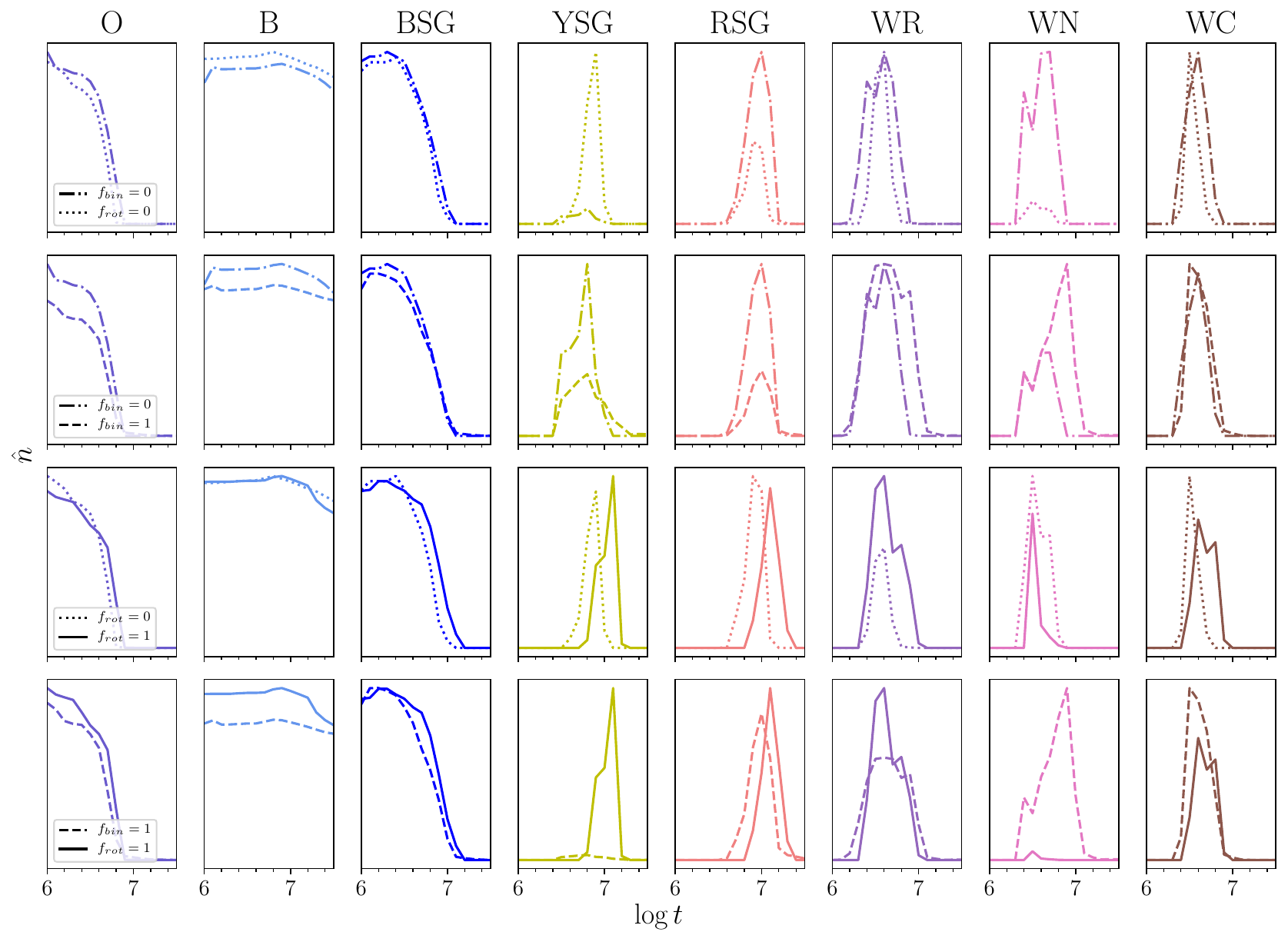}
\caption{Number of various stellar subtypes at $Z=0.014$ between $10^6$ and $10^{7.5}$ years. The top row compares the $f_{bin}=0$ population (dash-dotted) with the $f_{rot}=0$ population (dotted), the second row compares the $f_{bin}=1$ (dashed) and $f_{bin}=0$ (dash-dotted) populations, the third row compares the $f_{rot}=1$ (solid) and $f_{rot}=0$ (dotted) populations, and the bottom row compares the $f_{bin}=1$ (dashed) population with the $f_{rot}=1$ (solid) population. Both lines in a given panel are plotted on the same linear y-scale to allow for comparison.}\label{fig:counts_burst}
\end{figure*}

Photometric surveys of nearby massive stars \citep[e.g., the Local Group Galaxy Survey, LGGS; ][]{massey06,massey07}, can yield fairly complete catalogs after filtering for foreground contaminants \citep[e.g., ][]{massey09}, and follow-up narrow-band surveys can be used to find evolved emission line stars (e.g., \citealt{neugent11} and \citealt{neugent12,neugent18}). Even without follow-up spectroscopy, photometric measurements can then be used to categorize stars into broad spectral types. Thus it is useful to classify all timesteps of a given evolutionary track into one of a number of coarse spectral types, using the position of the model timestep on the HR diagram, as well as its surface composition, based on the model parameters listed in Table \ref{tab:mod_params}; we largely adapt the classification scheme from \citet{eldridge17}, shown in Table \ref{tab:count_criteria}. \citet{georgy13} use a similar classification scheme, with slightly different temperature or composition thresholds. All labels refer explicitly to spectral types. WNH corresponds to Hydrogen-rich Wolf-Rayet stars; O$f$ stars are O stars with particularly strong winds and \ion{He}{2} emission \citep{brinchmann08}. The BSG, YSG, RSG, and WR numbers are computed by summing the numbers of the indicated species, and applying the relevant luminosity threshold, as described below.

Where they are different, we adopt the Geneva criteria to classify the Geneva tracks, and use the BPASS criteria in cases where the BPASS classification is more specific than the Geneva classification (e.g., BPASS distinguishes between K and M stars). This choice is motivated by two facts. Firstly, in some cases, these criteria are used within the individual evolutionary tracks to distinguish between different prescriptions for, e.g., mass-loss. Secondly, the coupling between the outermost layer of a model star and a model stellar atmosphere in order to produce a synthetic spectrum is non-trivial. Indeed, the criteria for classifying WR stars have nothing to do with the mass-loss rate, which might be observed from such a spectrum. In lieu of synthesizing a spectrum for each timestep of each model, we defer to the creators of each code in how to best classify their models. However, it should be noted that the exact choice of the values presented in Table \ref{tab:count_criteria} do not drastically affect the results \citep{meynet03}. When we attempted to classify the Geneva tracks using the BPASS criteria, we found little substantive changes, except the rotating Geneva models produce more WN stars (still less than half the WNs produced by BPASS).

We then use the classification, age, and weight of each track to find the frequency of each spectral type in Table \ref{tab:count_criteria} as a function of time, binned to 51 time bins that are logarithmically-spaced between $10^6$ and $10^{11}$ years in 0.1 dex increments as described in Paper I. We also assign each model to one of 31 luminosity bins with 0.1 dex width between $\log(L) = 3$ and 6. Thus we can apply coarse luminosity thresholds to mimic observational completeness limits. This allows us to account for unresolved binaries by making the assumption that all stars below a given luminosity, including secondaries, are not detected, while all stars above this threshold are. While a somewhat simplistic assumption, the evolved stages of massive star evolution are so short-lived that most evolved massive stars have main sequence companions \citep[e.g.,][]{neugent18b,neugent19}. Of course, binaries with two evolved components do exist (e.g., WR+WR binaries), but are usually detectable via their wind interactions. In the case that spectral observations are of insufficient SNR or resolution to classify the components of such a binary, the WC/WN ratio would be unreliable. We encourage observational efforts dedicated to making a census of the massive component of stellar populations to discuss their insensitivity to unresolved binaries.

Each column in Figure \ref{fig:counts_burst} shows the number, $\hat{n}$, of the spectral types in Table \ref{tab:count_criteria}. For clarity, values on the y-axis are not shown, but both lines in a given panel are plotted on the same, linear scale. Line styles are used to indicate the four different populations, with dash-dotted lines for $f_{bin}=0$, dashed for $f_{bin}=1$, dotted for $f_{rot}=0$, and solid for $f_{rot}=1$. The top row corresponds to the single, nonrotating populations from both evolutionary codes, the second row shows the $f_{bin}=0$ and $f_{bin}=1$ populations, the third row shows the $f_{rot}=0$ and $f_{rot}=1$ populations, and the bottom row shows the $f_{bin}=1$ and $f_{rot}=1$ populations. Note that, as in Paper I, we do not include Luminous Blue Variables in our analysis. This is due to the present uncertainty in the evolutionary status of LBVs \citep{smith15,humphreys16,aadland18}, and the lack of a clear consensus in how to observationally classify a statistically significant number of them without long-term monitoring.

\subsection{Diagnostic Ratios}\label{subsec:diagnosticratios}

\begin{figure*}[!ht]
\centering
\includegraphics[width=\textwidth]{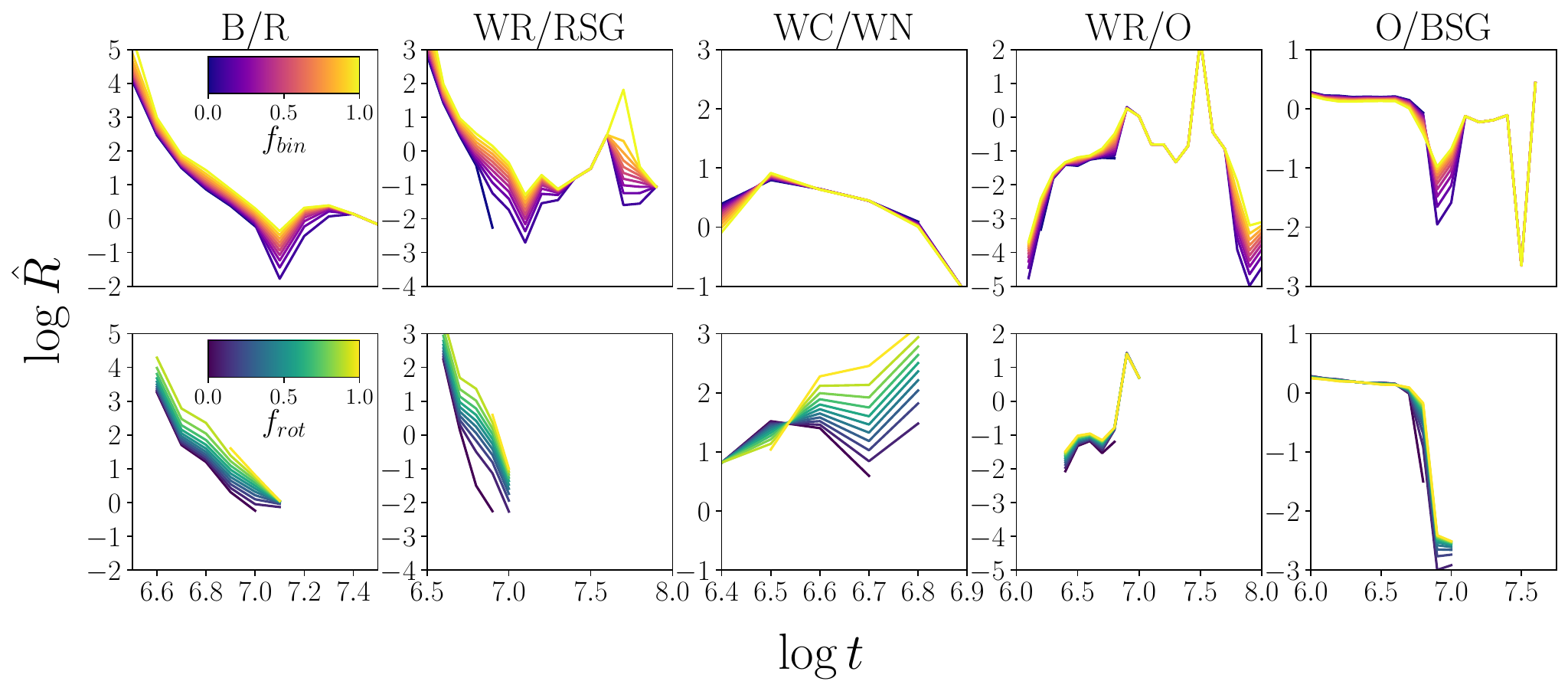}
\caption{Predicted values of five diagnostic ratios (where $B/R$ is shorthand for the ratio of blue to red supergiants), at solar metallicity, for the BPASS (top) and Geneva (bottom) populations as a function of time, for values of $f$ between 0 and 1, as indicated by the colorbar.}\label{fig:ratios_burst}
\end{figure*}

Because the calculated number counts are the frequency of each subtype per $10^6 M_\odot$ of stars formed, direct application to observed populations requires an estimate of the stellar mass of a population, $M_*$. Such measurements are often model dependent, and are based on inferences of the sometimes-undetected low-mass end of the population. Instead, ratios of the frequency of these types (hereafter ``number count ratios'') are independent of the stellar mass, while remaining sensitive to both rotation and binary interactions. 

We first construct the predicted number counts for subtypes in a population with a given $f_{bin}$ or $f_{rot}$ (generically $f$ hereafter). All notation used is summarized in Table \ref{tab:mod_params}. We calculate $\hat{n}_S$, the frequency of a subtype S at time $t$ and metallicity $Z$ as
\begin{equation}
\hat{n}_S(t,f,Z) = f \hat{n}_{S_{f=1}}(t,Z) + (1-f)\hat{n}_{S_{f=0}}(t,Z)
\end{equation}
where $\hat{n}_{S_{f=0}}$ and $\hat{n}_{S_{f=1}}$ are the frequencies in the $f=0$ and $f=1$ populations respectively. We note that, for the BPASS populations, the naive interpretation of $f_{bin}$ is rather straightforward: $f_{bin}$ is the fraction of binary stars in the population\footnote{There is a secondary factor here, in that $f_{bin}$ is mass-dependent \citep{duchene13}. However, all of the evolutionary phases here are mostly descended from O and early B stars (at least in the single-star paradigm), for which observed samples are too small to determine any mass dependence.}. However, for the Geneva populations, $f_{rot}$ would then correspond to the fraction of stars in the population born with $v_{ini}/v_{crit} = 0.4$ (i.e., rapid rotators), while realistic stars rotate with a range of velocities, from nonrotating to almost critically rotating; another interpretation might then be for $f_{rot}$ to correspond to the average initial rotation rate (i.e., $\langle v_{ini}/v_{crit} \rangle = 0.4 f_{rot}$, see, for example, \citealt{levesque12}). Ultimately, in both cases, $f$ is simply a factor used to linearly combine the {\it output number counts} from each population, and does not necessarily correspond to the fraction of binary/rotating star models that are used, a fact that the reader should be aware of when interpreting our results.

We then calculate number count ratios using
\begin{equation}
    \hat{R}_{S_1/S_2}(t,f,Z) = \hat{n}_{S_1}(t,f,Z)/\hat{n}_{S_2}(t,f,z)
\end{equation}
for two subtypes S$_1$ and S$_2$. 

In Paper I we described four number count ratios frequently found in the literature, as well as a novel number count ratio, O/BSG. Here we briefly describe these ratios, and the physical effects that they probe. We note that, in theory, our model populations would allow us to perform a search for the ratios and completeness limits that would best differentiate between different channels of stellar evolution. However the existing data tends to focus only on individual species, making the available space of ratios that can be measured quite small. In future work, we plan to perform this search in order to guide observers to evolutionary species for which an accurate census is most useful for constraining stellar evolution.

\begin{itemize}
    \item {\it B/R}: The ratio of the number of BSGs to RSGs (B/R) is among the most frequently used ratios, and has a long history as a metallicity diagnostic \citep{walker64,vandenbergh68,langer95,massey03AR}. It is sensitive to the physics governing the timescale of rightward evolution of stars on the HR diagram --- rotational/convective mixing --- as well as the interruption of a star's expansion by RLOF.
    \item {\it WR/RSG}: In the single star paradigm, the WR/RSG ratio probes the boundary between stars that experience only redward evolution and stars that lose enough mass to evolve blueward at the end of their lives \citep{conti83}, and is therefore also sensitive to metallicity \citep{maeder80}. Mass loss via binary channels serves to artificially boost this ratio by decreasing the number of RSGs, and commensurately increasing the number of WRs.
    \item {\it WC/WN}: The WC/WN ratio probes the evolution of stars that have already lost enough mass to become WRs. Thus it is mostly insensitive to the binary fraction, and is a very sensitive diagnostic of radiative mass loss in the WR phase \citep{vanbeveren80,hellings81}. 
    \item {\it WR/O}: The WR/O ratio probes the largest swatch of the mass spectrum considered here. As the only ratio in the literature with main sequence components, it is the most subject to contamination by unresolved O+O binaries \citep{maeder91} that is difficult to address via our simple implementation of completeness limits. 
    \item {\it O/BSG}: In Paper I, we introduced the O/BSG ratio. Both species are recovered by photometric censuses of bright blue stars in stellar populations, and don't require narrow band imaging or spectroscopic follow-up to detect. While, in theory, it is mostly sensitive to the main sequence lifetime (and thus rotational and convective mixing), some main sequence O stars are luminous enough to be classified as BSGs in our scheme. As we demonstrated, O/BSG is mostly insensitive to the binary fraction, except in a narrow window around $\log{t}=7$. This is due to lower mass stars losing enough of their envelopes via RLOF to evolve blueward, without losing enough Hydrogen to be labelled as WR stars --- observationally these might be classified as sdB/O stars, which are not counted separately from O and B stars in our classification scheme (see Section 3.2 and Figure 10 of \citealt{eldridge17}). This effect boosts the O/BSG ratio relative to the single star population until all of the O stars have evolved. Thereafter, only small numbers of models in the binary population reside within the Luminosity-Temperature-Composition boundary of O stars or BSGs at different times, causing rapid changes to O/BSG (for a particularly drastic example of a star rapidly entering and leaving the O and BSG regimes, see Figure 1 of \citealt{dornwallenstein18}).  
\end{itemize}

Figure \ref{fig:ratios_burst} shows the values of five different ratios vs. time for solar metallicity BPASS (top row) and Geneva (bottom row) populations with $0 \leq f \leq 1$ as indicated by the colorbar. Panels in the same column have identical bounds on the vertical axis for comparison. The bounds of the time axis have been chosen to highlight the time range during which each ratio is most dependent on $f$.

\section{Results}\label{sec:results}
\subsection{Comparing Rotating and Binary Model Populations}\label{subsec:comparing}

The differences between the single/nonrotating O and B stars in the top row of Figure \ref{fig:counts_burst} are minimal, where the dotted line shows the nonrotating Geneva population, and the dash-dotted line shows the BPASS single-star population. We warn the reader that, in general, the BPASS user manual cautions against using the $f_{bin}=0$ population in isolation; only the $f_{bin}>0$ populations can be considered reliable. However, we show the $f_{bin}=0$ results for completeness, and discuss the differences between the two $f=0$ populations in detail here and throughout the text.

Significant differences arise in the yellow supergiant phase, where the Geneva models predict the existence of far more YSGs. However, theoretical uncertainty in this very short-lived phase has long stymied our understanding of massive star evolution \citep{kippenhahn90}; for this reasons we do not use the YSG phase in our subsequent ratio diagnostics and caution against using it as a diagnostic of stellar population properties until it is better understood. The single BPASS models produce approximately twice as many RSGs as the nonrotating Geneva models. This is largely due to the fact that the BPASS models cross the HR diagram quicker (reflected in the significantly smaller number of YSGs compared to the Geneva models), increasing the amount of time the stars spend as RSGs before ending their lives. However, the two model sets also adopt slightly different mass loss prescriptions during the RSG phase: the Geneva tracks use mass loss rates from \citet{reimers75,reimers77} for the models less massive than 12 $M_\odot$ during the RSG phase, and a combination of mass loss rates from \citet{dejager88}, \citet{sylvester98}, and \citet{vanloon99} for more massive models, while BPASS only uses rates from \citet{dejager88} that are higher on average. This serves to modulate the increased numbers of RSGs seen in the single BPASS models. This difference in mass loss rates carries over into the WR phase, where BPASS produces far more H-deficient WN stars, while the Geneva models form more H-rich WNH stars (not shown). However, the overall numbers of WR stars (and WC stars) are similar between the two codes.

The second and third rows illustrate the various effects of binary interactions and rotation. The effects of both rotation and binarity can be seen increasingly clearly from the least to most evolved phases. In particular, RLOF decreases the number of YSGs and RSGs. This causes a boost in the number of WR stars in the $f_{bin}=1$ population at late times. Rotation prolongs the length of the early evolutionary phases, serving to delay the onset of the RSG and YSG phases. Rotating models also produce higher mass loss rates thanks to luminosity-dependent mass loss prescriptions and the higher luminosities of rotating stars, beginning at the terminal age main sequence and persisting through their post-main-sequence evolution \citep{maeder00b,ekstrom12}, as well as contributions from mechanical mass loss (e.g., mass loss via the stellar equator from matter rotating above the critical velocity). We also see an increase in the number of WR stars formed in the rotating Geneva populations; this boost primarily manifests as an increase of WNH stars, with slight {\it decreases} in the number of WN and WC stars. This is the result of a longer lifetime for the WNH phase and subsequent shorter lifetimes for the WN and WC phase (a consequence of rotational mixing), as well as efficient mass loss producing lower-mass WNH stars (for more discussion of rotation effects of WR subtypes see \citealt{georgy12}). Finally, the bottom panel serves as a comparison between the binary BPASS population (dashed line) and the rotating Geneva population (solid line) as shown in the above rows. In summary: rotation causes a delay in the appearance of evolved supergiants and an increase in the number of WNH stars, while binary interactions effectively trade evolved supergiants for WR stars. However, the exact timing and degree of these effects as a function of initial mass/luminosity is more complicated, and this effect manifests in the observed stellar populations as we will demonstrate. 

The predictions of the number count ratios in Figure \ref{fig:ratios_burst} are fairly similar between the Geneva and BPASS populations at $\log{t}\lesssim6.5-7$. Both binary interactions and rotation introduce similar effects, especially in B/R and WR/RSG. The Geneva models predict a higher overall value of WC/WN before $\log{t}\sim6.75$ (due to the increase/decrease of WNH/WN stars respectively in the Geneva models), and a higher local maximum of WR/O at $\log{t}\sim6.8$. At this time, the WR component of the Geneva populations becomes increasingly dominated by WC and WNH stars at increasing $f_{rot}$, while the BPASS WRs are largely WN type, which is reflected in the multiple orders of magnitude difference in the prediction for WC/WN. At increasingly later times, more stars in binaries become stripped. Models that lose enough Hydrogen increase the number of WRs well after $\log{t}\sim6.8$, when the Geneva populations predict the last WRs have died, while models with only moderate mass transfer/loss become O stars/BSGs, depending on their luminosity --- this is reflected in the rapid changes in the O/BSG ratio at late times. Overall, depending on the ratio chosen and the approximate age of the population being analyzed, different ratios are most sensitive to age, $f_{bin}$, $f_{rot}$ or all three. For example, WC/WN is incredibly sensitive to $f_{rot}$ in moderately evolved populations, while B/R is a relatively powerful age indicator in the earliest populations.

Below solar metallicity, the dominant differences are that all WR stars in the Geneva populations are H-rich (WNH in our labelling scheme), and binary interactions become increasingly important for producing WRs in the BPASS populations. The former is consistent with \citet{georgy13} and \citet{groh19}, and is a known feature of the Geneva models \citep{leitherer14}. While this might indicate that all H-deficient WRs at low metallicity are formed by binary interactions, other possibilities and evolutionary pathways exist, which we defer to work focused more specifically on WR populations.

\subsection{Comparisons with Real Data}\label{subsec:data}
\subsubsection{Ensuring Self-Consistency} 

Two important effects must be considered before directly comparing some observed number count ratio to a theoretical prediction of this ratio, to ensure that the quantities being compared are identical: 

\begin{itemize}
    \item The value (and corresponding uncertainty) reported by an observer must be an estimate of the {\it underlying} number count ratio, $\hat{R}$ (an intrinsic characteristic of the stellar population belonging to the set of real numbers), rather than the raw observed ratio, $R$ (which is a characteristic of the data belonging to the set of rational numbers due to the integer nature of the measurement).
    \item The theoretical population must approximate the observed population, and reflect the completeness of the catalog of massive stars (which may vary with spectral type).
\end{itemize}

We first consider how to estimate number count ratios and confidence intervals from observed data. Here we present a novel framework for making point-estimates (with corresponding uncertainties) for an {\it intrinsic} number count ratio. Much like photons in the low-count regime, the frequency of finding a given spectral type is determined by Poisson statistics (this is especially true for counting massive stars, where ``shot noise'' is the dominant source of uncertainty). In particular, the measurement uncertainty of the frequency of two subtypes, $n_1$ and $n_2$, can be approximated as $\sigma_{n_1} = \sqrt{n_1}$ and $\sigma_{n_2} = \sqrt{n_2}$ respectively. Past studies that calculate the number count ratio \citep[e.g.,][]{massey03,neugent12}, apply traditional propagation of uncertainties, and report the {\it observed} ratio $R=n_1/n_2$, with corresponding uncertainty $\sigma_{R} = R\sqrt{n_1^{-1} + n_2^{-1}}$. As discussed by \citet{neugent12}, this approach is problematic, and more sophisticated corrections can be made. However, an additional problem exists in that, with few enough stars (e.g., the Wolf-Rayet population of the SMC, where 1 WC star is known; \citealt{neugent18}), or where the true underlying ratio is large, yet finite, $n_2=0$ is well-within a ``3$\sigma$'' error bar, and a measurement of $R=\infty$ could have been made. 

This is a well-studied problem in the X-ray astronomy community, with a tractable solution within a Bayesian framework. \citet{park06} derive the posterior probability distribution of colors and hardness ratios for X-ray sources, in the limit of few (or no) photon counts. The problem here corresponds to a special case where the background is guaranteed to be 0, which simplifies the calculations somewhat; unknown sample contamination by foreground stars can be accounted for with minimal added complexity.

Say we measure a ratio $R=n_1/n_2$. Both $n_1$ and $n_2$ are assumed to be Poisson variables with (unknown) expectation values $\hat{n}_1$ and $\hat{n}_2$. What we wish to report is an estimate of the {\it true} ratio, $\hat{R} \equiv \hat{n}_1/\hat{n}_1$, which is a property of the underlying stellar population --- indeed, it is the exact quantity plotted in Figure \ref{fig:ratios_burst}. From Bayes' theorem, the probability distribution for $\hat{n}_1$ given a measurement of $n_1$ is given by

\begin{equation}\label{eq:bayes}
p(\hat{n}_1|n_1) \propto p(\hat{n}_1)p(n_1|\hat{n}_1)
\end{equation}

and similarly for $p(\hat{n}_2|n_2)$, where $p(\hat{n}_1)$ reflects our prior knowledge on the value of $\hat{n}_1$, and $p(n_1|\hat{n}_1)$ is the likelihood of drawing $n_1$ from a Poisson distribution with expectation value $\hat{n}_1$. For a prior, we adopt $p(\hat{n}_1) \propto \hat{n}_1^{\phi - 1}$. As discussed in \citet{park06} and \citet{vandyk01}, this is a special case of a $\gamma$-prior:

\begin{equation}
p(\hat{n}_1,\alpha,\beta) = \frac{1}{\Gamma(\alpha)}\beta^{\alpha}\hat{n}_1^{\alpha-1}e^{-\beta\hat{n}_1}
\end{equation}

with $\alpha = \phi$ and $\beta \rightarrow 0$. This choice of prior ensures that the posterior probability function takes the same parametric form as the likelihood function. \citet{park06} found that, in Monte Carlo simulations, the choice of $\phi$ only has a moderate impact on the coverage (the percentage of simulations where the ground truth value of $\hat{R}$ is within a 95\% confidence interval). We choose $\phi = 1/2$, which generally provides the best coverage for the observed number counts reported in typical extragalactic surveys.

Assuming $\hat{n}_1$ and $\hat{n}_2$ are independent (i.e., no stars of type 1 would also be counted as type 2\footnote{In one example below, $n_1$ is subset of $n_2$. There, a simple transformation can be made, but handling more complex situations is nontrivial.}), the joint posterior distribution is $p(\hat{n}_1,\hat{n}_2|n_1,n_2) = p(\hat{n}_1|n_1)p(\hat{n}_2|n_2)$. Transforming $\hat{n}_1 = \hat{R}\hat{n}_2$ and marginalizing over $\hat{n}_2$, 
\begin{equation}
p(\hat{R}|n_1,n_2)d\hat{R} = d\hat{R}\int_{\hat{n}_2} d\hat{n}_2 \hat{n}_2 p(\hat{R}\hat{n}_2,\hat{n}_2|n_1,n_2)
\end{equation}

Utilizing Eq. \eqref{eq:bayes}, and substituting in the prior and likelihood functions, 
\begin{equation}\label{eq:posterior}
p(\hat{R}|n_1,n_2) \propto  \int_{\hat{n}_2} d\hat{n}_2 \hat{R}^{\phi - 1} \hat{n}_2^{2\phi - 1} \frac{\hat{R}^{n_1}\hat{n}_2^{(n_1+n_2)}e^{-\hat{n}_2(\hat{R}+1)}}{n_1!n_2!}
\end{equation}
We use {\tt emcee} \citep{foremanmackey13}, a Markov Chain Monte Carlo package, to sample the joint posterior probability distribution for $\hat{R}$ and $\hat{n}_2$ with 100 walkers initialized around the observed value $R$, 500 burn-in steps that are discarded, and an additional 3000 steps to explore the stationary distribution of walkers. In cases with $R \rightarrow \infty$ or $R\rightarrow0$, we force $R$ to be in the range $[10^{-10},10^5]$ when initializing the walkers. We estimate the value of both $\hat{R}$ and $\hat{n}_2$, as well as a 68\% (1$\sigma$) confidence interval, using the 16th-, 50th-, and 84th-percentile values of the samples.\footnote{Software for performing these calculations, as well as reproducing all of the results in this work, is available online at \url{https://github.com/tzdwi/Diagnostics/}} The key advantages of this method are that the estimated quantity can be directly compared to the model predictions, that the reported errorbars correspond to the actual posterior probability distribution (and can be assymetric), and that the estimate of $\hat{R}$ is meaningful even if $n_1$ or $n_2$ are 0.

The challenge of accounting for complete samples is discussed in depth in Paper I. Here we reiterate that accounting for incompleteness in the observed samples --- here defined as the lowest luminosity to which all stars of a given subtype have been found --- is critical, and incorrectly handling or ignoring this effect can result in biases of $\sim0.1$ dex in the estimated population age, and lead to incorrect results. In order to account for this effect, we include a luminosity cutoff $L_{\rm cut}$, that can be tuned for each subtype under consideration, and does not include models with $L < L_{\rm cut}$. Thus, even if the sample is assembled spectroscopically, or has a limiting magnitude in some photometric band that corresponds to different luminosity thresholds depending on the effective temperatures of the different subtypes, the model populations can be adjusted accordingly.

We note that assuming an observed sample is 100\% complete above $L_{cut}$, and no stars are detected below it is a somewhat simplistic assumption. Below we compare our models to actual observed samples. In the two star clusters that we focus on, the data mostly come from focused studies that are designed to detect a given species. These stars are the brightest objects in a given part of the color magnitude diagram, have been followed up spectroscopically to remove contaminants, and the sample should be complete above the lowest luminosity star. In the Magellanic Clouds, the current existing samples of WRs and RSGs claim to be mostly-complete censuses of both species down to quite stringent magnitude limits \citep{neugent18,neugent12}. Some confusion arises in the counting of BSGs; however, the topic of bright blue stars in the Magellanic Clouds is currently being debated. We defer here to authors with more expertise \citep{aadland18}. Finally, observational bias, particularly in spectroscopic searches for WR stars, is likely to lead to missed weak-lined WR stars or WR stars with less evolved companions when not carefully accounted for, as discussed by \citet{neugent18}.

\subsubsection{Starburst Comparisons} 
We now wish to test our models in an environment where both sets of populations produce roughly identical predictions in a simple stellar population. From Figure \ref{fig:ratios_burst}, the best examples are young ($<10$ Myr) star clusters, where we can assume that all of the stars belong to a single burst of star formation \citep[see caveats in][]{gossage18}. At these young ages, most WRs formed by binary interactions are evolved from progenitors that were massive enough to become WRs anyway. There are very few such clusters with enough confirmed members to adequately sample the IMF. With a mass of $M_*\approx5\times10^4$ $M_\odot$ cluster \citep{andersen17}, and a well-studied cohort of evolved massive stars \citep{clark05,crowther06}, including a large number of BSGs, and an appreciable amount of WRs and RSGs, Westerlund 1 (Wd 1) is perhaps the best Galactic test bench for our model populations. In Paper I, we demonstrated that, when including binary effects and accounting for completeness, we can use two number count ratios, to estimate an age consistent with \citet{crowther06}, who use a single diagnostic ratio and did not account for binarity or completeness. We can now apply our updated proscription for estimating number count ratios, as well as the rotating populations.

\begin{figure}[!ht]
\plotone{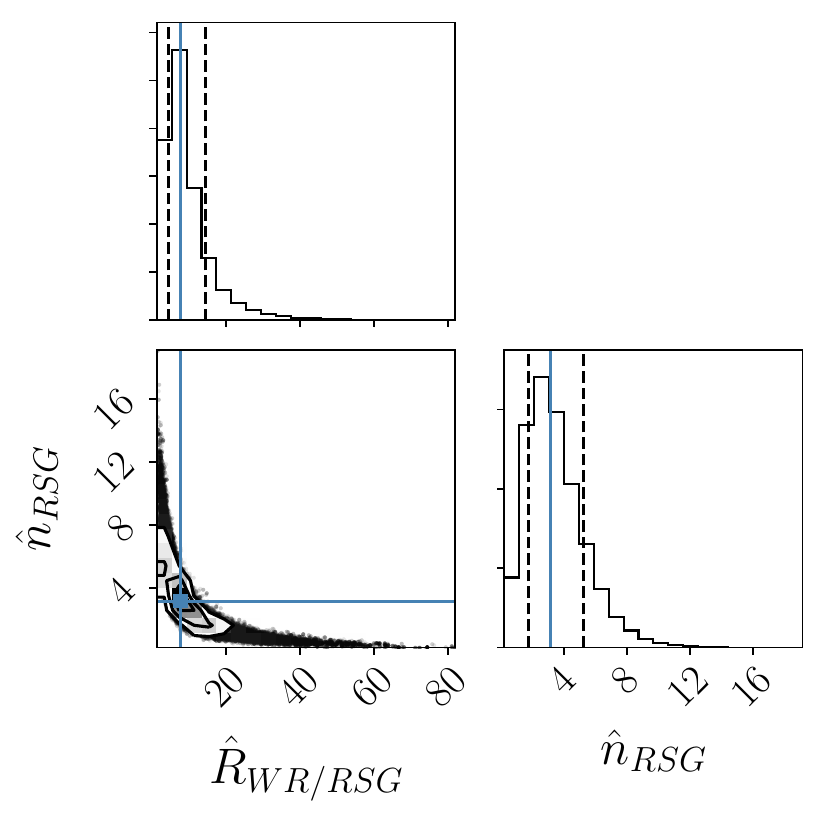}
\caption{Posterior distribution samples of the estimate of WR/RSG for Wd 1. The 1-D histograms show the marginalized posterior distribution for the true ratio $\hat{R}$, and the true number of RSGs, $\hat{n}_{RSG}$, as well as the point estimates (in blue vertical lines) and 68\% confidence intervals (in dashed vertical lines)}\label{fig:corner}
\end{figure}

We first apply the Monte Carlo method described above to estimate the value of two ratios, O/BSG and WR/RSG. Using data from \citet{clark05} and \citet{crowther06}, we count $n_O = 22$, $n_{BSG} = 29$, $n_{WR} = 24$, and $n_{RSG} = 3$. As an illustration, the samples of the joint posterior distribution for $\hat{R}_{WR/RSG}$ and $\hat{n}_{RSG}$ are shown in the bottom-left panel of Figure \ref{fig:corner}, along with the marginalized posterior distributions for each parameter and accompanying point estimates (solid blue vertical line) and 68\% confidence intervals (dashed black vertical lines). Note that because $n_{RSG}$ is so low, the distribution of $\hat{R}_{WR/RSG}$ is very skewed, and the 97.5th-percentile upper limit is much higher than the reported 84th-percentile. All of the O stars in our sample are also blue supergiants, and so our assumption of independent variables no longer holds. Instead, we estimate $\hat{R}_{O/(BSG-O)}$ where BSG-O refers to all BSGs that are not O stars. We then transform each posterior sample of O/(BSG-O) into a sample of O/BSG estimates. With this method we measure intrinsic values of $\hat{R}_WR/RSG = 7.619 ^{+6.748}_{-3.194}$ and $\hat{R}_{O/BSG} = 0.755 ^{+0.072}_{-0.084}$.

\begin{figure*}[ht!]
\centering 
\leavevmode 
\includegraphics[width={0.329\linewidth}]{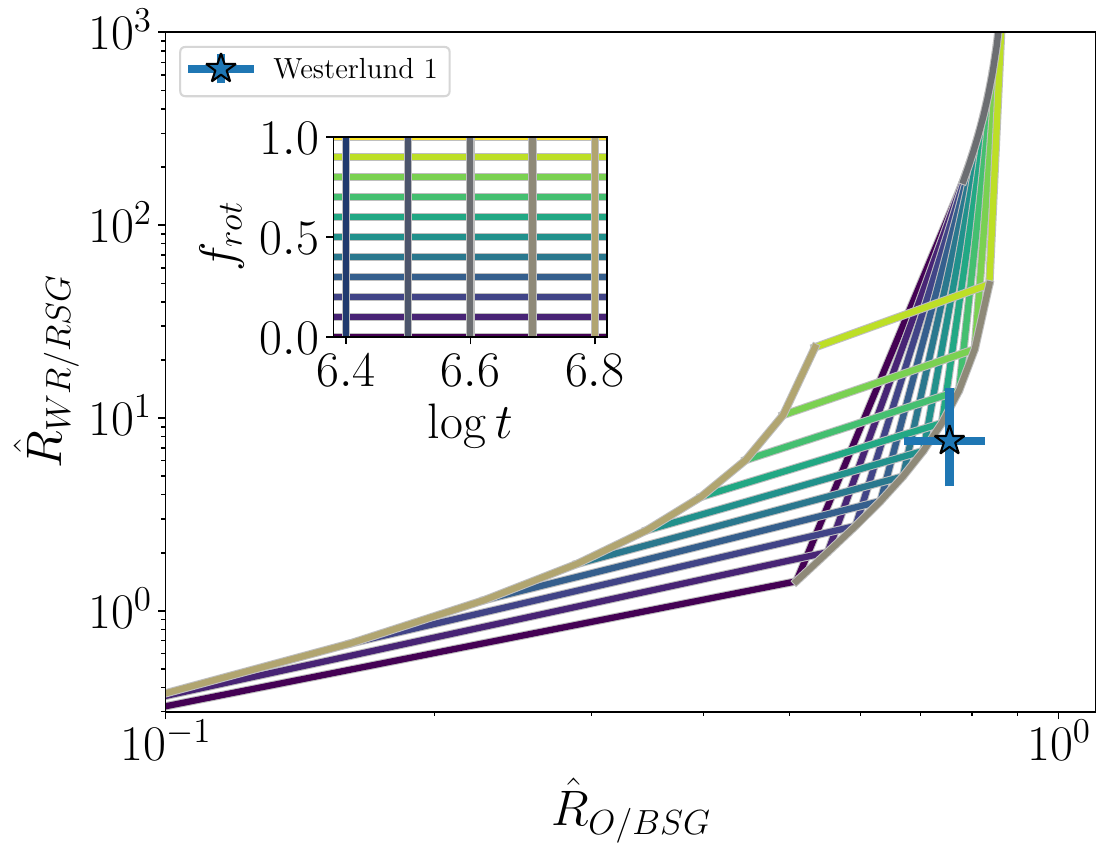}
\hfil 
\includegraphics[width={0.329\linewidth}]{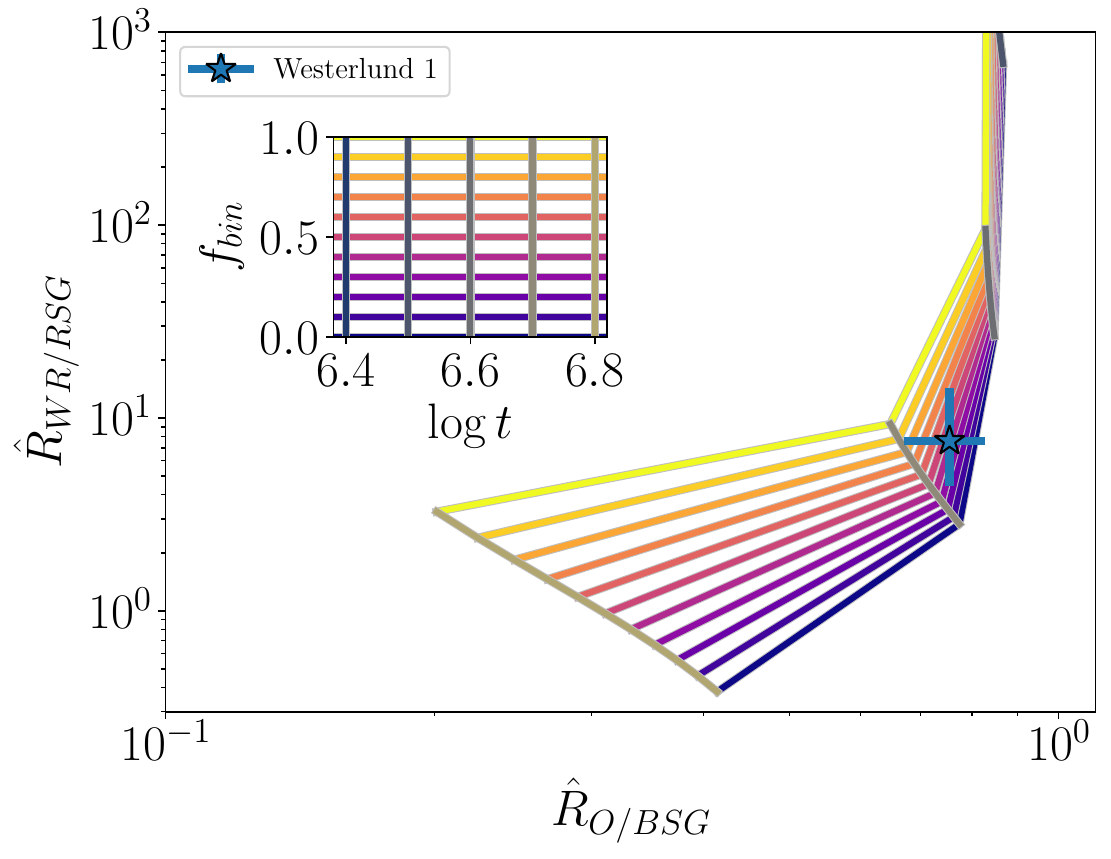}
\hfil
\includegraphics[width={0.329\linewidth}]{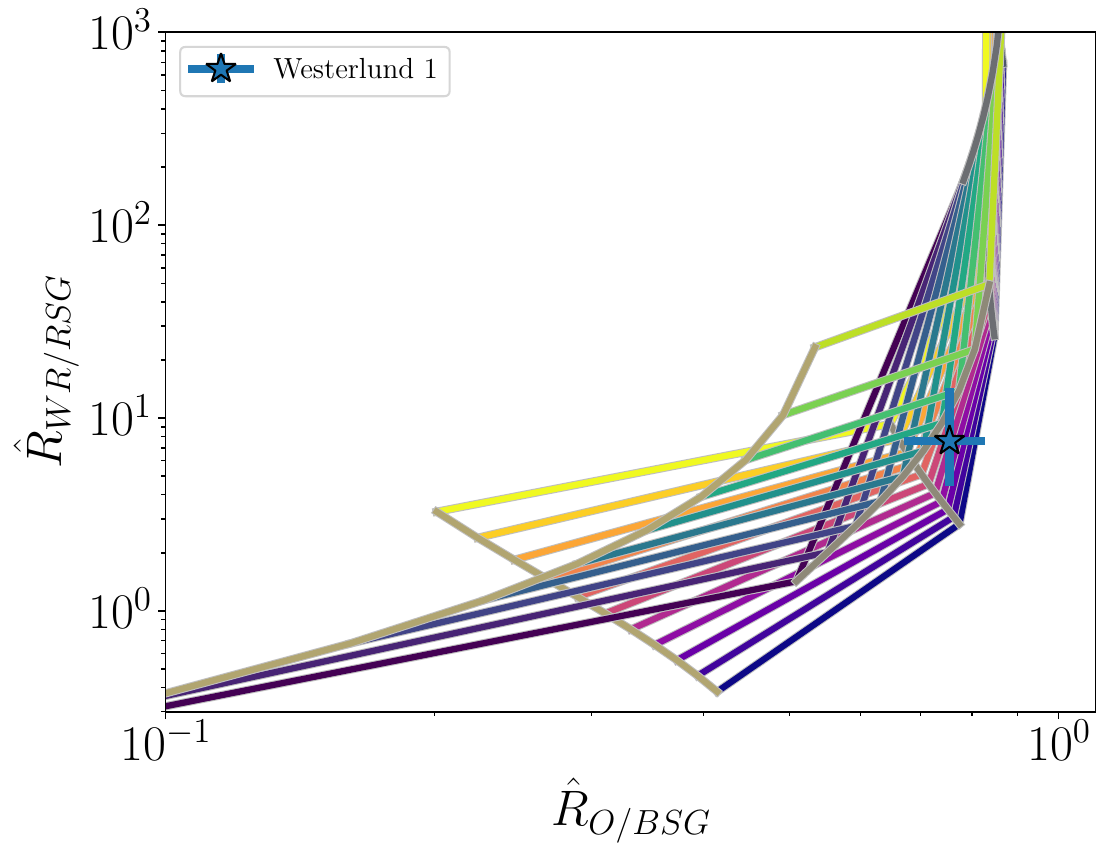}
\caption{$\hat{R}_{WR/RSG}$ vs $\hat{R}_{O/BSG}$ for the Geneva (left) and BPASS (center) populations, calculated on a grid of $6.4 \leq \log{t} \leq 6.9$, and $0 \leq f \leq 1$, as shown by the inset panels. The rightmost panel shows both grids overlain on top of each other, with identical color coding. An estimate for $\hat{R}_{WR/RSG}$ and $\hat{R}_{O/BSG}$ in Westerlund 1, as well as corresponding 68\% confidence intervals, is calculated using data from \citet{clark05} and \citet{crowther06} and shown in blue.}\label{fig:wd1_example}
\end{figure*}

We now wish to estimate age and $f$ from the data. Figure \ref{fig:wd1_example} shows the predictions for WR/RSG vs. O/BSG, calculated on a grid of Geneva (left) and BPASS (center) populations with varying age (with $6.4 \leq \log{t} \leq 6.9$) and $f$. Lines of constant age and $f$ are shown; the inset plot can be used to translate from the ratio-space into $f$ and log age. The models incorporate completeness limits consistent with the lowest luminosities of each subtype reported by \citet{clark05} and \citet{crowther06} (specificially, $L_{\rm cut} = 4.9,4.9,5.1,4.9$ for O stars, BSGs, WRs, and RSGs respectively). The right panel shows both grids overlain on top of each other. Combined, the two sets of models predict more-or-less identical values of both ratios as a function of age, especially for the $f=0$ populations (shown in purple in all three panels). The data for Wd 1, plotted in blue, are consistent with an age of $\log{t}=6.7$ (or approximately 5 Myr) using both model sets. This example is useful for validating our models, in that we demonstrate that the data are consistent with both model sets, as expected. Interestingly, the lines of constant age in the two grids are nearly orthogonal, implying that, with higher signal to noise data, these ratios can yield a measurement of both $f_{bin}$ and $f_{rot}$.

\begin{figure*}[ht!]
\centering 
\leavevmode 
\includegraphics[width={0.329\linewidth}]{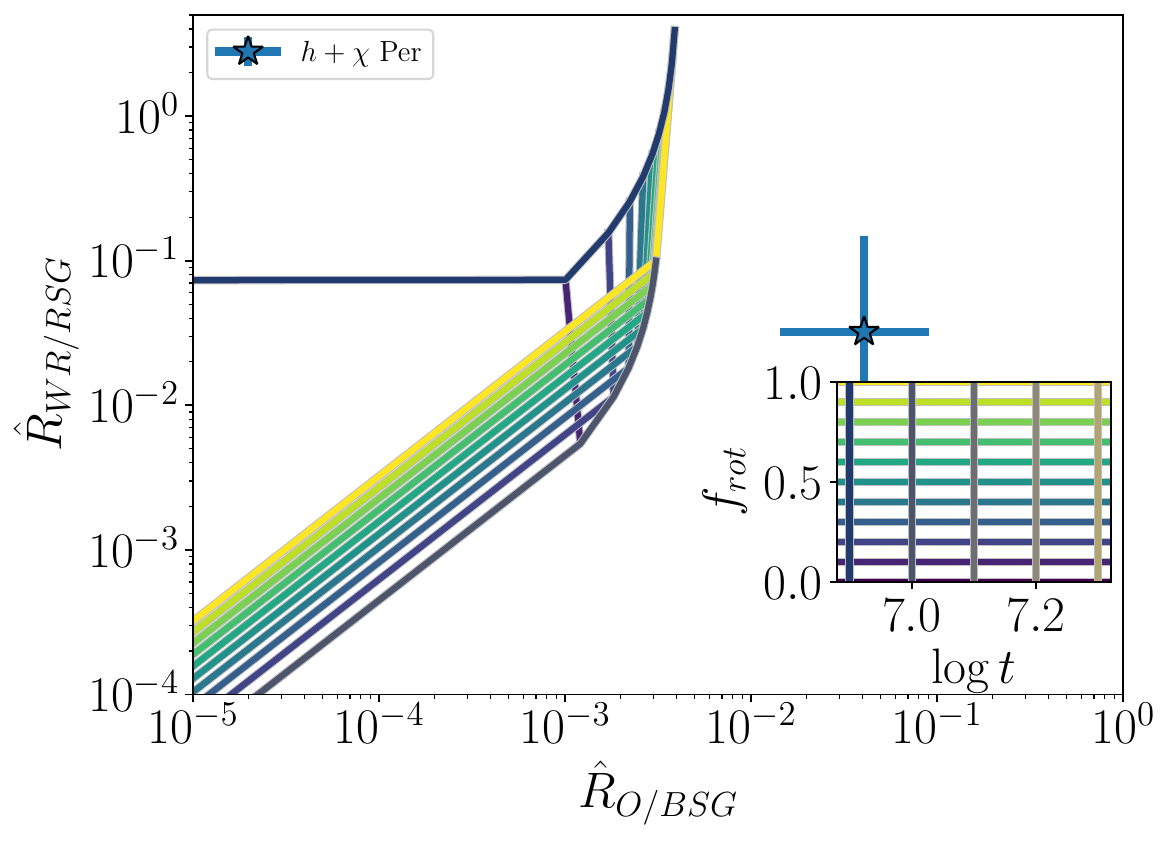}
\hfil 
\includegraphics[width={0.329\linewidth}]{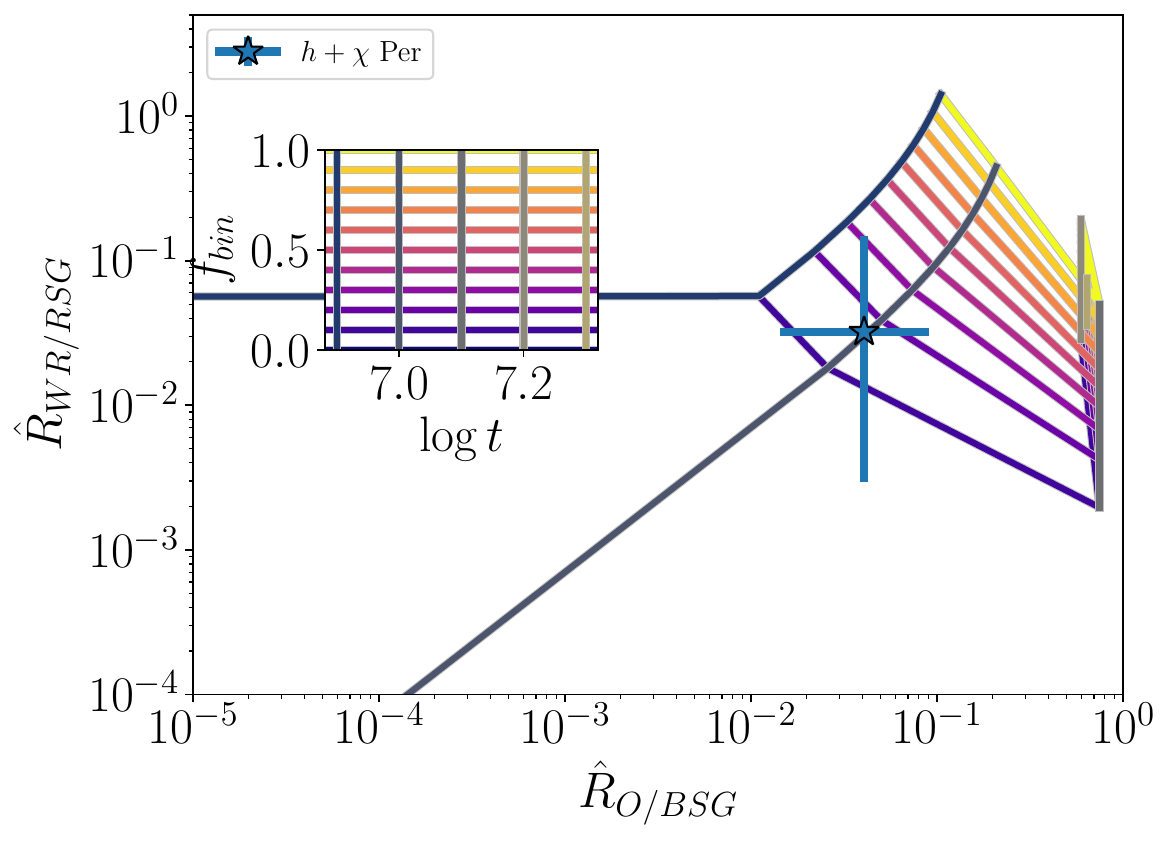}
\hfil
\includegraphics[width={0.329\linewidth}]{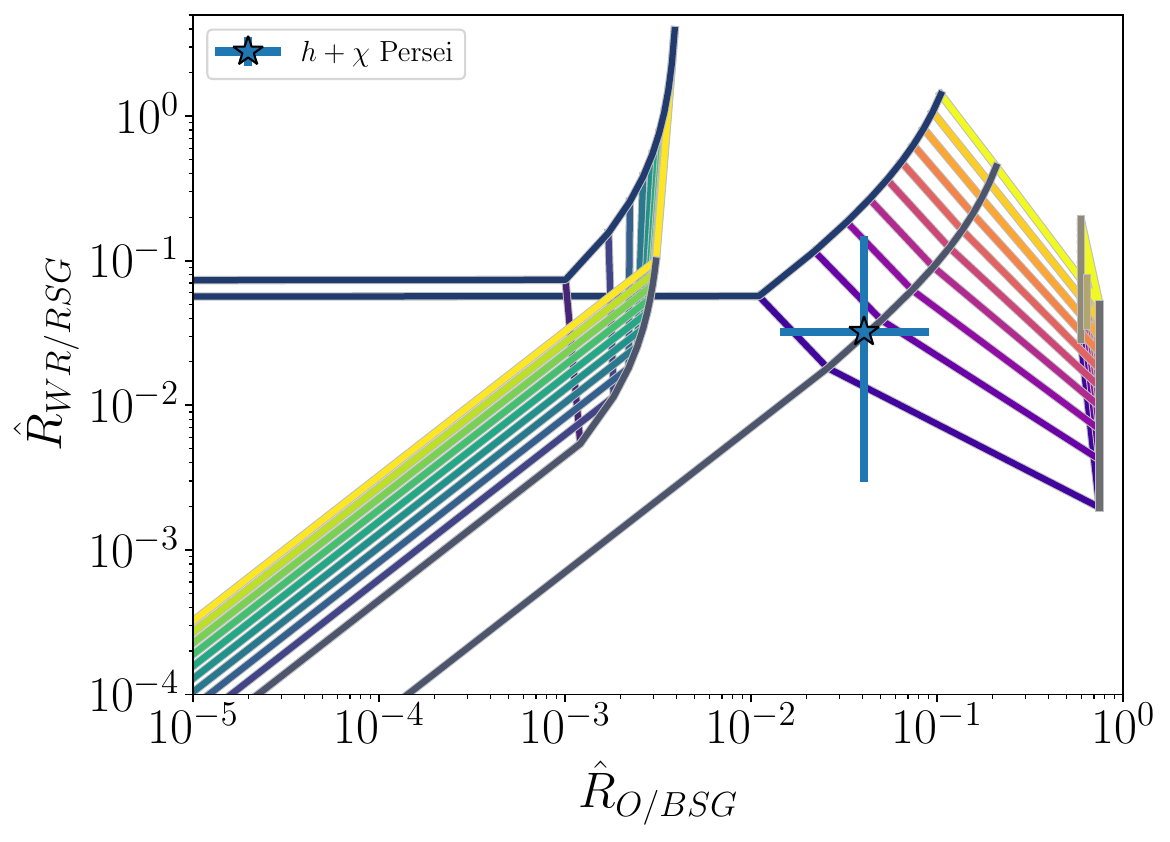}
\caption{Similar to Figure \ref{fig:wd1_example}: $\hat{R}_{WR/RSG}$ vs $\hat{R}_{O/BSG}$ for the Geneva (left) and BPASS (center) populations, now calculated on a grid of $6.9 \leq \log{t} \leq 7.4$. An estimate for $\hat{R}_{WR/RSG}$ and $\hat{R}_{O/BSG}$ in $h+\chi$ Persei, as well as corresponding 68\% confidence intervals, is calculated using data from \citet{currie10} and shown in blue.}\label{fig:hchiper_example}
\end{figure*}

In the example above focused on Wd 1, most of the stars in the sample were OB dwarfs or supergiants, where the differences between binary and rotating scenarios are small. For intermediate age clusters ($\log{t} \gtrsim 7$), more stars are expected to be found in increasingly evolved states, and the model grids no longer overlap. Figure \ref{fig:hchiper_example} is similar to Figure \ref{fig:wd1_example}, but calculated for SSPs with log-ages between 6.9 and 7.4 (approximately 8 and 25 Myr), with a luminosity threshold of $\log{L} = 4.9$ for all subtypes but O stars (comparable with the sample in \citealt{currie10}, to which we compare below). The two model grids are separated by orders of magnitude in O/BSG, and showcase quite different behavior as a function of age in both ratios. This is due to the WR stars in the Geneva models dying, while increasing amounts of primary stars in the BPASS models are being stripped by binary interactions. These models boost the value of both ratios, depending on their luminosity and whether they lose enough Hydrogen to be classifed as WRs; if not, they tend to instead be classified as O stars. Note that we cannot visually compare the $f=0$ case from both model sets, where $\hat{R}_{WR/RSG}\rightarrow0$. Both grids reflect this, as they asymptote off the bottom left of the plot.

There exists only one stellar population that is massive enough to test our synthetic populations in this age regime: $h+\chi$ Persei. Photometric and spectroscopic studies of the members of $h+\chi$ Per have determined ages of 13-14 Myr ($\log(t)\approx7.1$) for both clusters \citep{slesnick02,currie10}. While these studies have revealed a population of O stars, BSGs, and RSGs, no obvious WRs have been found, and thus we were unable to compare the data with the models in Paper I (see that work for a detailed discussion of detecting low-mass WRs). Now, we can use the data from Table 3 in \citet{currie10} to count 1 O star (HD 14434), 29 BSGs, and 7 RSGs, and estimate $\hat{R}_WR/RSG = 0.032 ^{+0.115}_{0.029}$ and $\hat{R}_O/BSG = 0.041 ^{+0.050}_{0.026}$, assuming 0 observed WRs. These values are plotted in blue in Figure \ref{fig:hchiper_example}. Note that the samples O stars and BSGs are {\it independent}, and we do not need to perform the same transformation as above. Because \citet{currie10} study the main sequence down to G0 dwarfs, the luminosity cutoffs applied here are consistent with the data. The point estimates for both ratios are consistent with the BPASS models, though the Geneva models are not completely excluded. Interestingly, this difference is driven primarily by our measurement of O/BSG, and not WR/RSG (which we estimate without observing any WR stars). Using the BPASS model grid, the data corresponds to an age of $\sim10$ Myr, consistent with previous age estimates. 

\subsubsection{Constant Star Formation Histories}

While these simple stellar populations in our Galaxy are useful, the 68\% confidence intervals on our number count ratio estimates are too wide to accurately determine age and $f_{bin}$/$f_{rot}$. To obtain a larger sample of evolved massive stars, we turn to galaxies in the Local Group. Not only do these galaxies contain more massive stars, they also sample a broad range of metallicities. However, these populations have complex star formation histories (SFHs), which we describe in Paper I.

\begin{figure}[!ht]
\includegraphics[width={1.1\linewidth}]{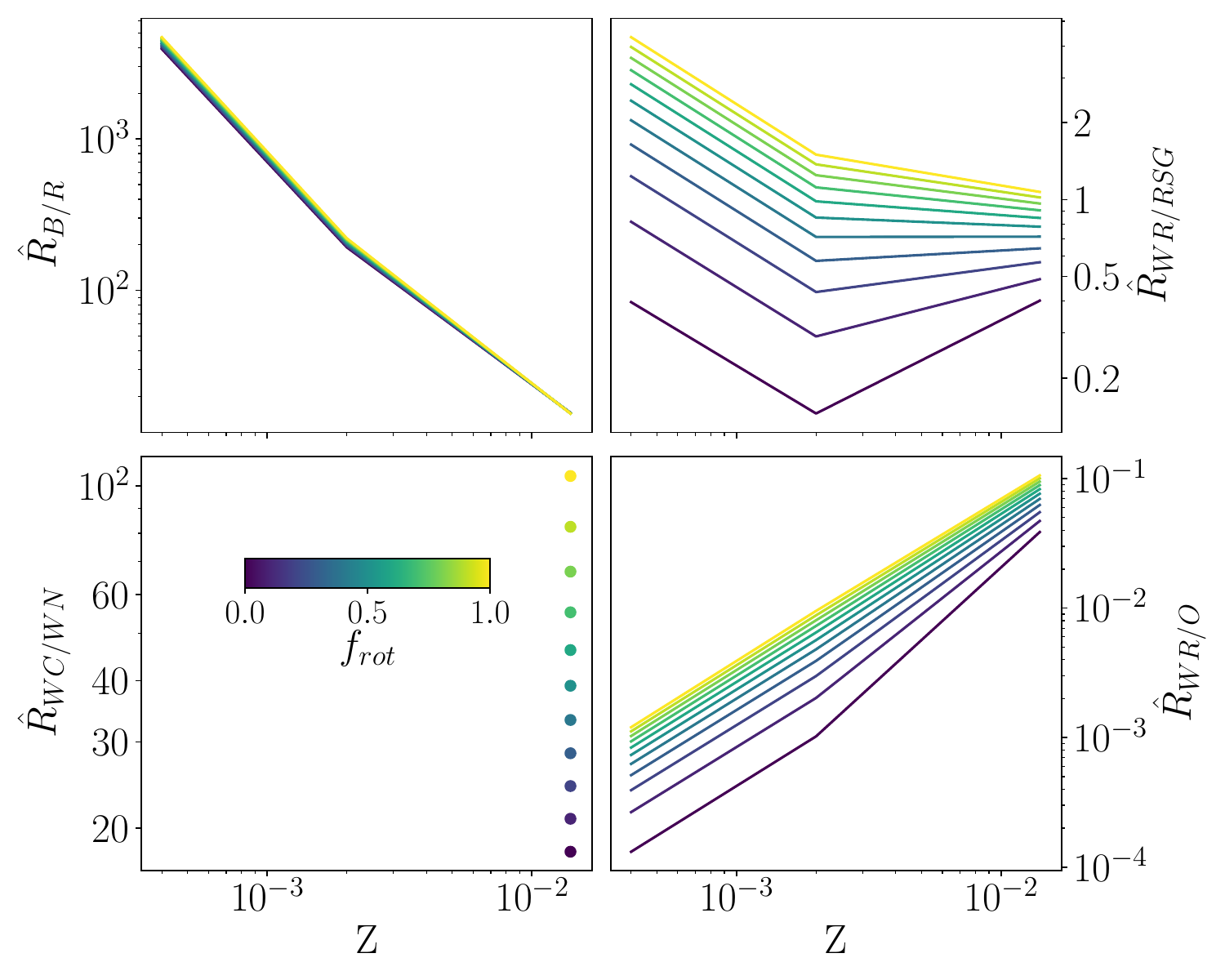}
%\plotone{ratios_constSFH.pdf}
\caption{Predicted values of B/R (top left), WR/RSG (top right), WC/WN (bottom left), and WR/O (bottom right) for Geneva populations with constant star formation, and values of $f_{rot}$ between 0 and 1, as indicated by the color bar. A minimum luminosity of $\log(L) = 4.9$ is assumed for all subtypes.}\label{fig:ratios_constSFH}
\end{figure}

Here, we only consider populations that are constantly forming stars, following \citet{eldridge17}. Despite including stars from all age bins, all of the evolved types of massive stars considered here are only sensitive to, at most, 50-100 Myr of star formation, after which the populations of massive stars reach an equilibrium. Figure \ref{fig:ratios_constSFH} shows WR/O, B/R, WR/RSG, and WC/WN as a function of metallicity and $f_{rot}$ for galaxies constantly forming stars. A minimum luminosity of $\log(L) = 4.9$ is applied to be consistent with typical extragalactic studies (i.e., where only the supergiant population is complete). As discussed above, the Geneva models don't produce WC or WN stars below solar metallicity; the predictions at $Z_\odot$ are presented instead as individual points. Consistent with past models and existing observations \citep{vandenbergh73,humphreys79,maeder80}, B/R is the least sensitive ratio to $f_{rot}$, and is a good indicator of metallicity, while WR/RSG is largely independent of metallicity \citep[a fact that has long been in tension with observations, see section 6.4 of ][and citations therein]{levesque17}.

\begin{figure*}[ht!]
\centering 
\leavevmode 
\includegraphics[width={0.329\linewidth}]{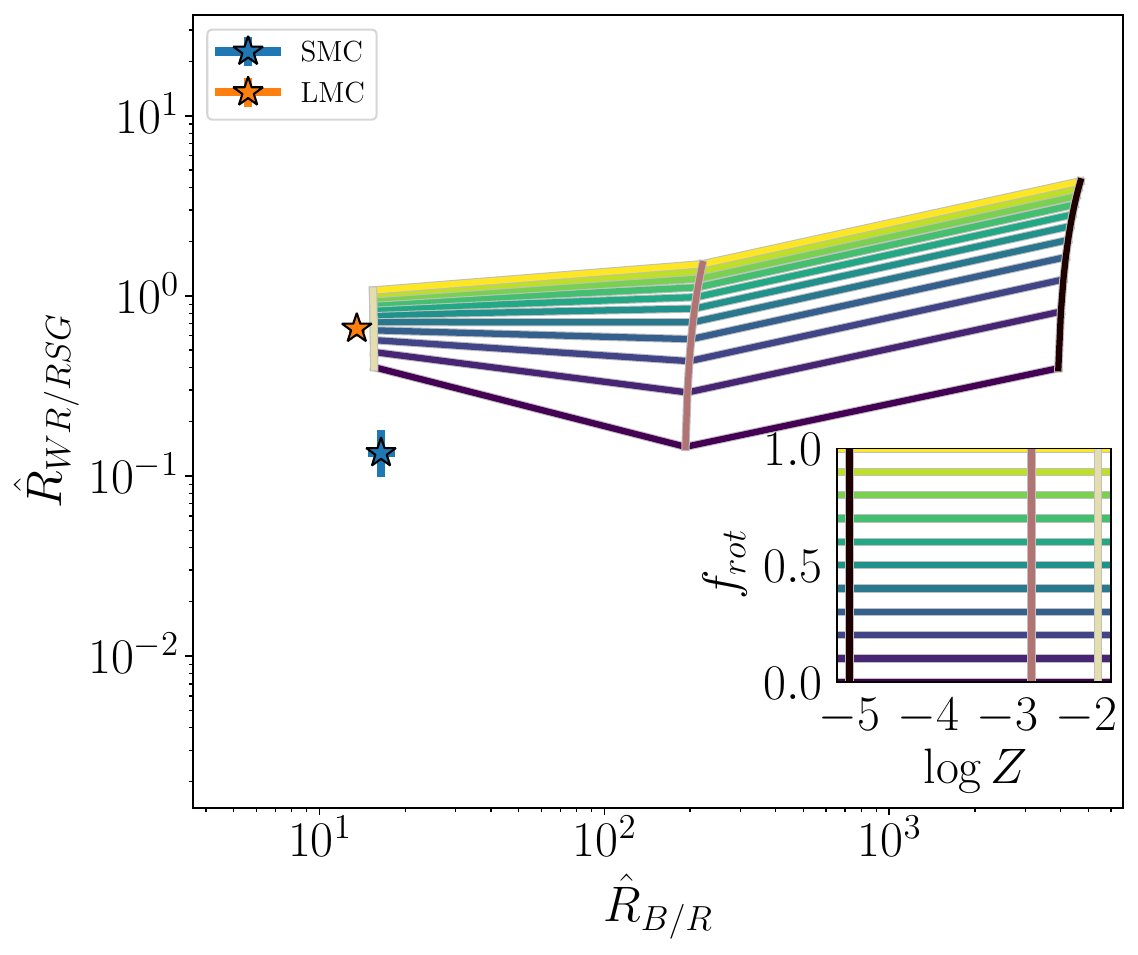}
\hfil 
\includegraphics[width={0.329\linewidth}]{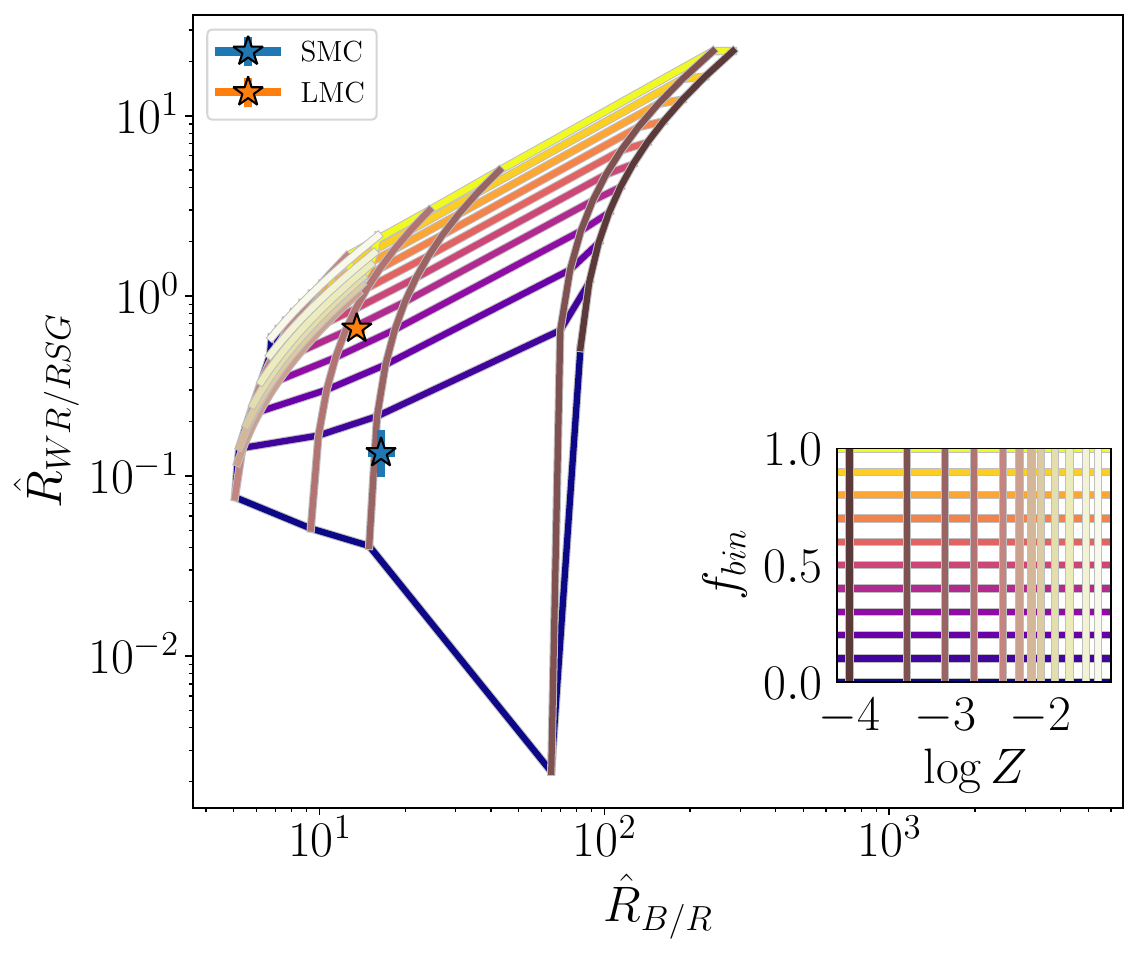}
\hfil
\includegraphics[width={0.329\linewidth}]{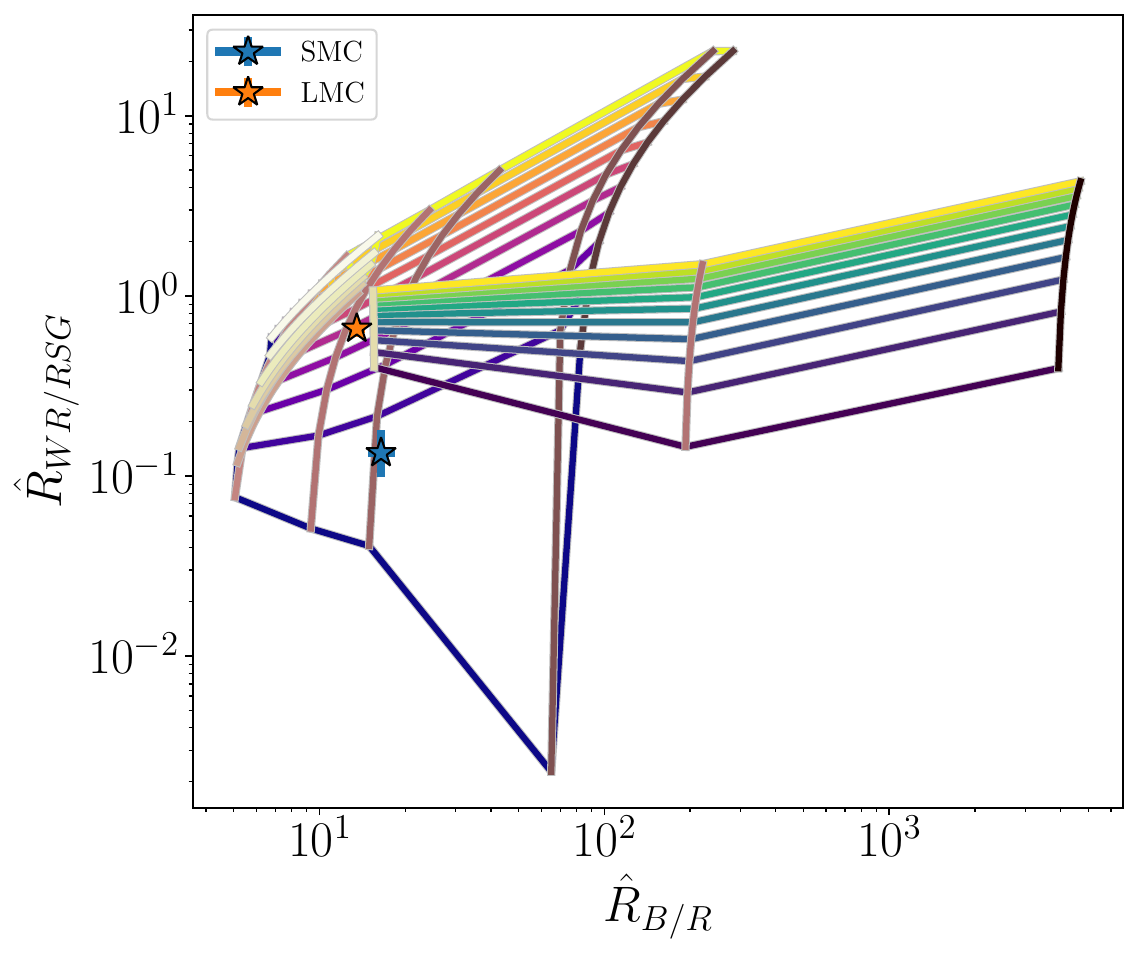}
\caption{Similar to Figure \ref{fig:wd1_example}: $\hat{R}_{WR/RSG}$ vs $\hat{R}_{B/R}$ for the Geneva (left) and BPASS (center) populations, assuming constant star formation, and calculated on a grid of $10^{-5} \leq Z \leq 0.04$. Estimates for $\hat{R}_{WR/RSG}$ and $\hat{R}_{B/R}$ in the LMC and SMC, as well as corresponding 68\% confidence intervals, are shown in orange and blue, respectively. Data are from \citet{massey03}, \citet{neugent11}, and \citet{neugent12}.}\label{fig:MC_example}
\end{figure*}

Recent measurements of the frequency of all three subtypes are only available for the Magellanic Clouds (MCs). Figure \ref{fig:MC_example} shows $\hat{R}_{WR/RSG}$ vs. $\hat{R}_{B/R}$ using the Geneva (left), and BPASS models (center), with both grids overlaid (right), with lines of constant $f_{rot}$, $f_{bin}$ and $Z$, as indicated with the inset grids. We note that the $f=0$ case from both grids do not agree; BPASS predicts approximately an order of magnitude smaller values in both ratios. This is largely due to the significant difference in the number of RSGs predicted (see the top row of Figure \ref{fig:counts_burst}), and thus, the difference in the mass loss prescription used by each code. However, both grids reside in largely the same part of this ratio space. Furthermore, consistent with Figure \ref{fig:ratios_constSFH}, B/R remains a useful proxy for metallicity, and WR/RSG is a useful diagnostic of either $f_{bin}$ or $f_{rot}$.

For both the Large Magellanic Cloud (LMC, plotted in orange) and Small Magellanic Cloud (SMC, plotted in blue), we estimate $\hat{R}_{B/R}$ using data from \citet{massey03} and $\hat{R}_{WR/RSG}$ from \citet{massey03}, \citet{neugent11}, and \citet{neugent12}, and find $\hat{R}_{B/R} = 13.521 ^{+0.937}_{0.865}$, $\hat{R}_{WR/RSG} = 0.659 ^{+0.071}_{0.064}$ in the LMC, and $\hat{R}_{B/R} = 16.449 ^{+1.898}_{1.658}$, $\hat{R}_{WR/RSG} = 0.134 ^{+0.046}_{0.036}$ in the SMC. The Geneva models fail to reproduce the observations, while the BPASS models underpredict the metallicity of both clouds ($Z\approx0.001/0.002$ for the SMC/LMC respectively). In both cases, the data indicate that the initial $f_{bin}$ or $f_{rot}$ is larger in the higher metallicity LMC than in the lower metallicity SMC, consistent with the results in \citet{dornwallenstein18}. We stress here that these results refer to the {\it natal} values of these parameters. Indeed, the exact opposite trend is predicted and observed in the frequency of X-ray binaries \citep{belczynski04}; however, this is a reflection the metallicity-dependent angular momentum evolution of binary systems. Measurements of the frequency of O star binaries (which in theory have not undergone mass transfer interactions in our model populations, and should serve as a decent proxy for the initial value of $f_{bin}$) reveal that $f_{bin}$ does indeed increase from the LMC ($f_{bin}\sim0.5$, \citealt{sana13}) to the Galaxy ($f_{bin}\sim0.7$, \citealt{sana12}). Meanwhile, the observed rotation rates of metal-poor stars are expected to be higher (i.e., $f_{rot}$ decreases with metallicity), due to stars being more compact at low $Z$ \citep{chiappini06}. However, the agreement between the overall trends in the data and those predicted by both model sets imply that number count ratios may be the best way to infer $f_{bin}$ or $f_{rot}$ in galaxies where direct measurements are currently infeasible.

\section{Discussion and Conclusion}\label{sec:conclusion}

Before discussing our results, we note two important caveats. Our results here assume that both stellar evolution codes accurately describe the evolution of the modelled stars. If this were true, then the $f=0$ cases from each population should be identical, as both populations contain only single, nonrotating stars. Of course, both stellar evolution codes make different assumptions; while we refer the reader to the papers describing both codes for more details, we note that the different mass loss prescriptions in particular can result in the differing numbers of predicted cool supergiants. 

We are furthermore assuming that both codes are completely accurate descriptions of rotation or binary interactions. Indeed, both codes are now seen as the industry standard for modelling their respective effects in massive stars. However, both codes do exhibit shortcomings. For example, the Geneva group only provides models that rotate at $0.4 v_{crit}$. More rapidly rotating models are available, but only in a limited mass window below 15 $M_\odot$. With rapidly rotating models at higher mass incorporated into our model populations, we would be able to probe stars with significantly enhanced rotational mixing and enhanced mass loss --- perhaps increasing the number of WNs while decreasing the number of WNHs. 

Meanwhile, the BPASS team is the only group that explicitly models RLOF in binary systems and makes their results publicly available. However, as noted in \citet{eldridge17}, BPASS makes a number of simplifying assumptions in its treatment of circular orbits, rotation, and common envelope evolution. Furthermore, BPASS does not model systems with initial orbital periods shorter than one day. Regardless, BPASS is still successful at producing all classes of observed binary systems; in theory, some merger products are not modelled, but those products are either the result of very short-period systems --- which merge very early, and evolve as single stars (J. Eldridge 2018, private communication), which may make result in an effectively top-heavier IMF --- or are incredibly rare objects (e.g., Thorne-\.Zytkow Objects, \citealt{thorne75,thorne77}).

These caveats aside, it is still important to examine how these two widely-used codes compare to each other and to observations when predicting the evolution and populations of massive stars, and to consider this comparison when interpreting the use and application of these models in future work. Our main results are as follows:

\begin{enumerate}
    \item While rotation and binary interactions predict qualitatively similar effects on stellar populations, the predictions for the detailed makeup of simulated simple and complex stellar populations show dramatic differences between the two evolutionary scenarios. While some of these can be attributed to fundamental differences in the Geneva and BPASS codes (as discussed in Section \ref{sec:themodels}), it is also clear that rotation and binarity have quantitatively different effects on the evolution of massive star populations (in particular, the diagnostic number count ratios discussed in Section \ref{subsec:comparing}, and shown in Figure \ref{fig:ratios_burst}).
    \item We introduced a novel Bayesian method of estimating both the value and error of diagnostic ratios in the low-number count regime typical of samples of massive stars. This method, combined with our implementation of completeness limits in the model populations, make our comparisons between models and data more accurate.
    \item The data from observed Galactic populations agree with the grids of simulated SSPs, but suffer from poor signal-to-noise. The increased sample sizes in Local Group galaxies allows us to make higher-precision estimates of diagnostic ratios, and show that measurements of the natal binary fraction or rotation rate of stellar populations beyond the Magellanic Clouds may be possible with currently obtainable data. In the coming decades, JWST and WFIRST are scheduled to launch, giving us immediate access to red/optical-mid IR photometry, as well as sparsely sampled lightcurves for massive stars well-beyond the Local Group. Using comparable existing data in the Galaxy and Magellanic Clouds, it is possible to accurately fit for coarse spectral types akin to those used here (Dorn-Wallenstein et al. in prep).
    \item Figures \ref{fig:wd1_example}, \ref{fig:hchiper_example}, and \ref{fig:MC_example} show a comparison between intrinsic number count ratios derived from grids of model stellar populations and inferred from observed data. For the starburst populations, we show that the ages consistent with the observed data are reasonably close to already published age estimates after correcting for completeness. However, the true power of this technique lies in its ability to derive the values and uncertainties of unknown parameters that are critically important for stellar evolution --- e.g. $f_{bin}$ or $f_{rot}$ --- in environments where traditional means of measuring these quantities are expensive or impossible given current technology. In this case, deriving the likelihood of obtaining a given set of star count ratios given a set of values for these parameters is nontrivial. Future work will focus on applying Approximate Bayesian Computation (ABC, \citealt{sunnaker13}) to derive constraints on $f_{bin}$ and compare them to existing values found in the literature \citep{sana12,sana14}.
    \item Finally, we stress that rotating stars are found in binary (and higher order) systems. A complete model of stellar evolution should not designate one or the other effect as secondary. Rather, as we demonstrate here, binary interactions and rotation and produce both similar and contradictory effects in stellar populations, and a rigorous simultaneous treatment of both is necessary.
\end{enumerate}

\acknowledgments
TZDW acknowledges J. Eldridge and C. Georgy for their advice and help. This research  was  supported  by  NSF  grant AST 1714285 awarded to EML. This work made use of v2.2.1 of the Binary Population and Spectral Synthesis (BPASS) models as described in \citet{eldridge17} and \citet{stanway18}.

This work made use of the following software:

\vspace{5mm}

\software{Astropy v3.2.2 \citep{astropy13,astropy18}, h5py v2.9.0, Matplotlib v3.1.1 \citep{Hunter:2007}, makecite \citep{makecite18}, NumPy v1.17.2 \citep{numpy:2011}, Python 3.7.4, IDL 8.3}

\bibliography{DL_bib}
\bibliographystyle{aasjournal}

\end{document}